\documentclass[11pt,a4paper,dvipdfmx]{article}
\usepackage{heppub}
\usepackage{amsmath, bm}
\usepackage{graphicx}

\title{Bi-Local Fields in AdS${}_5$ Spacetime}

\author[a]{Kenichi Aouda,}
\author[a]{Shigefumi Naka,}
\affiliation[a]{Department of Physics, College of Science and Technology, Nihon University, Tokyo 101-8308, Japan}
\author[b]{and Haruki Toyoda}
\affiliation[b]{Junior College, Funabashi Campus, Nihon University, Funabashi 274-8501, Japan}

\emailAdd{naka@phys.cst.nihon-u.ac.jp}

\abstract{ Recently, the bi-local fields attract the interest in studying the duality between $O(N)$ vector model and a higher-spin gauge theory in AdS spacetime. In those theories, the bi-local fields are realized as collective one's of the $O(N)$ vector fields, which are the source of higher-spin bulk fields. Historically, the bi-local fields are introduced as a candidate of non-local fields by Yukawa. Today, Yukawa's bi-local fields are understood from a viewpoint of relativistic two-particle bound systems, the bi-local systems. We study the relation between the bi-local collective fields out of higher-spin bulk fields and the fields out of the bi-local systems embedded in AdS${}_5$ spacetime with warped metric. It is shown that the effective spring constant of the bi-local system depends on the brane, on which the bi-local system is located. In particular, a bi-local system with vanishing spring constant, which is similar to the bi-local collective fields, can be realized on a low-energy IR brane.

 }

\keywords{Higher Spin Gravity, AdS-CFT Correspondence, Field Theories in Higher Dimensions}

\arxivnumber{1603.09542}

\begin{document}

\maketitle

\section{Introduction}

In the development of theories of massless higher-spin fields (HS), it is recognized that a cosmological constant of background spacetime is necessary to construct consistent theories of those fields\cite{Fradkin-Vasiliev-1,Fradkin-Vasiliev-2,Fradkin-Vasiliev-3}. In particular, higher-spin field theories under AdS backgrounds are expected as an important route studying AdS/CFT correspondence\cite{Polyakov-1,Polyakov-2}. According to this line of approach, there are many attempts to study AdS dual of conformal vector fields, which are sources of HS in AdS spacetime. It is also expected that those HS are realized in tensionless limit of string theory. In particular, recently, there arises an interesting point of view such that the aggregate of those HS is dual to a bi-local collective field out of conformal vector fields in the large $N$ limit. In terms of $O(N)$ vector fields $\phi^a(x),\,(a=1,2,\cdots,N)$, the bi-local collective field, there, is given by
\begin{align}
 \Phi(x,x^\prime)=\sum_{a=1}^N\phi^a(x)\phi^a(x^\prime).  \label{BL-CF}
\end{align}
Further, as a constructive approach,  the studies have also been made on an effective action of the bi-local collective fields\cite{Collective Fields-1,Collective Fields-2,Collective Fields-3,Collective Fields-4}, which provide Feynman diagrams associated with the CFT under some conditions.

Meanwhile, the bi-local fields have another history in the context of non-local field theories started by Yukawa at 1948\cite{Yukawa-1,Yukawa-2}, which are intended to introduce a physical constant with dimension of length to elementary particle theories under consideration of divergence problems and a variety of properties in elementary particles. In the beginning, Yukawa's bi-local fields have a meaning of two-particle systems constrained by a definite spacelike distance. Soon afterward, it was pointed out that such a two-particle system is reduced to free two particles by a canonical transformation\cite{Hara}. In response to this criticism, Yukawa modified his model so as to include an interaction between two particles. Nowadays, this type of bi-local field theories is discussed within the framework of relativistic two-particle systems bounded by a potential depending on spacelike distance of those particles. In this sense, such a modified bi-local system should be understood as a reduced model of relativistic string\cite{bi-local-1,bi-local-2}; and so, Yukawa's original attempt should be regarded as a bi-local counterpart of tensionless string.

The purpose of this paper is, thus, to investigate the bi-local systems in AdS${}_5$ spacetime by taking aim at the relation between Yukawa's bi-local field theories with vanishing spring constant and the bi-local collective fields in higher-spin field theories. In the next section, we try to construct a classical action of bi-local systems embedded in AdS${}_5$ spacetime. Therein, the two-body interaction in this curved spacetime is introduced by means of the geodesic interval connecting two particles. In section three, we discuss the bi-local field equation in bulk, the wave equation for the first quantized bi-local system in AdS${}_5$ spacetime. The analysis of the bi-local field, the one-particle wave function of the bi-local system, in respective branes is discussed in section four. It is shown that in particular, the bi-local fields in the low-energy IR brane are reduced to those of bi-local systems with vanishing spring constant due to an exponential hierarchy in energy scale. The section five is devoted to summary and discussion
\footnote{
. Some of those results were presented at \lq\lq CST \& MISC Joint Symposium on Particle Physics, 2015\rq\rq .
}.

\section{Embedding of a bi-local system in AdS${}_5$ spacetime}

The AdS${}_5$ spacetime with anti de Sitter radius $l$ is realized as the hyper-surface described by coordinates $(X^\mu,X^4,X^5)$ satisfying  $\eta_{AB}X^{A}X^{B}=\eta_{\mu\nu}X^{\mu}X^{\nu}+(X^4)^2-(X^{5})^2=-l^2$, where $(A)=(\mu,4,5)$ and $\mbox{diag}(\eta_{\mu\nu})=(-+++)$. The transformation $X^\mu=e^{-ky}x^\mu$, $X^4=l\sinh(-ky)-\frac{1}{2l}x^2e^{-ky}$, and $X^5=l\cosh(-ky)+\frac{1}{2l}x^2e^{-ky}$ with $k=l^{-1}$ can define another coordinate system of independent variables $(x^{\hat{\mu}})=(x^\mu,y)$. In this coordinate system, the spacetime can be characterized by the warped metric
\begin{align}
 \eta_{AB}dX^A dX^B=g_{\hat{\mu}\hat{\nu}}dx^{\hat{\mu}}dx^{\hat{\nu}}=e^{-2ky}\eta_{\mu\nu}dx^\mu dx^\nu+dy^2, \label{metric}
\end{align}
which is used in Randall-Sundrum model to address the Higgs Hierarchy Problem\cite{Randall-Sundrum-1,Randall-Sundrum-2}. In what follows, we consider the  AdS${}_5$ spacetime described by the coordinate system $(x^\mu,y),\,(0 \leq y \leq L)$ with this warped metric
\footnote{We set that the Planck energy scale brane and the low-energy brane are located respectively at $y=0$ and $y=L$; and, we regard $kL\simeq 32 \sim 50$ so that $e^{-kL}\simeq 10^{-14}\sim 10^{-22}$.}
.

Now, we set the action of a bi-local system in this curved spacetime so that
\footnote{In this paper, we use the unit $\hbar=c=1$.} \\
\begin{align}
S &=\int d\tau \frac{1}{2}\sum_{i=1}^2 \left\{e_{(i)}^{-1} g_{\hat{\mu}\hat{\nu}}D{x}_{(i)}^{\hat{\mu}} D{x}_{(i)}^{\hat{\nu}}-U\left(x_{(2)},x_{(1)}\right)e_{(i)} \right\},~ \left(~D{x}^{\hat{\mu}}_{(i)}=\frac{dx_{(i)}^{\hat{\mu}}}{d\tau}-(-1)^i\delta_y^{\hat{\mu}}\theta~\right) \nonumber \\
 &=\int d\tau \frac{1}{2}\sum_{i=1}^2 \left[ e_{(i)}^{-1}\left\{\eta_{\mu\nu}e^{-2ky_{(i)}}\dot{x}_{(i)}^\mu \dot{x}_{(i)}^\nu+\left(\dot{y}_{(i)}-(-1)^i\theta \right)\right\}^2 -U\left(x_{(2)},x_{(1)}\right)e_{(i)} \right],  \label{action}
\end{align}
where $\tau$ and $e_{(i)} \, (i=1,2)$ are respectively a time ordering parameter of dynamical variables and einbeins in $\tau$ space. The $\theta$ is an auxiliary variable, which transforms in the same way as $\dot{y}$ under the transformation of $\tau$. We have introduced the $\theta$ term to restrict the relative motion of $y$\cite{Abe}, although the covariance of this formalism is spoiled unless $\theta\rightarrow 0$.  The $U(x_{(2)},x_{(1)})$, $U_{2,1}$ simply, is a bi-scalar function representing the interaction of two particles at $x_{(i)},\,(i=1,2)$ with the same numerical value of $\tau$. According to a previous paper\cite{Schok-wave} we define this interaction term in such a way that
\begin{align}
 U(x_{(2)},x_{(1)})=2\kappa^2\sigma(x_{(2)},x_{(1)})+\omega , \label{potential}
\end{align}
where $\kappa$ and $\omega$ are positive constants with dimension of mass square; and, $\sigma(x_{(2)},x_{(1)})$ is the geodesic interval defined by\cite{deWitt}
\begin{figure}
\begin{center}
\includegraphics[width=5cm]{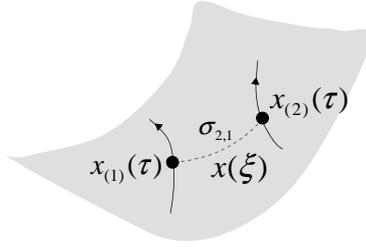}
\end{center}
\label{fig:1}
\caption{The figure shows the geodesic interval $\sigma_{2,1}$ connecting two particles in a curved spacetime so that $x_{(1)}(\tau)=x(\xi_0)$ and  $x_{(2)}(\tau)=x(\xi_1)$. The real lines are world lines of respective particles; and, the dotted line designates the geodesic having $x_{(i)},(i=1,2)$ as its ends.}
\end{figure}
\begin{align}
 \sigma(x_{(2)},x_{(1)})=(\xi_2-\xi_1)\int_{\gamma,\xi_1}^{\xi_2}d\xi{\cal L}\, ,~\left({\cal L}=\frac{1}{2}g_{\hat{\mu}\hat{\nu}}\frac{dx^{\hat{\mu}}}{d\xi}\frac{dx^{\hat{\nu}}}{d\xi} \right). \label{sigma}
\end{align}
The geodesic equation is equivalent to the Euler-Lagrange equation reading ${\cal L}$ as a Lagrangian.  Substituting the solution with conservative quantities along the geodesic for (\ref{sigma}), the geodesic interval is obtained as function of both ends of the geodesic (Fig.1); that is, $\sigma_{2,1}$ becomes a function of $x_{(i)}^{\hat{\mu}}(\tau),(i=1,2)$ only. 

 The $\sigma(x_{(2)},x_{(1)})$, $\sigma_{2,1}$ simply,  is  equal to one half the square of the distance along the geodesic between $x_{(1)}$ and $x_{(2)}$, which tends to $\frac{1}{2}\eta_{\hat{\mu}\hat{\nu}}( x_{(2)}-x_{(1)})^{\hat{\mu}}( x_{(2)}-x_{(1)})^{\hat{\nu}}$ according as $g_{\hat{\mu}\hat{\nu}}\rightarrow \eta_{\hat{\mu}\hat{\nu}}$. Thus, in such a flat spacetime limit, the $S$ in (\ref{action}) represents the action of a two-particle system bounded by a relativistic harmonic oscillator potential with a spring constant $\kappa^2$. 

Now, in a single-valued region of $y(\xi)$ such as $0\leq y(\xi) \leq L$ bounded by $y(\xi_0)=0$ and $y(\xi_L)=L$, the geodesic equations for $x^{\mu}(\xi)$ and $y(\xi)$ two kinds of constants along the geodesic $\gamma$ such that
\begin{align}
 V^{\mu} &=e^{-2ky(\xi)}x^{\prime\mu}(\xi), \label{constant velocity} \\
 K &=\frac{1}{2}\left(y^{\prime 2}(\xi)+e^{2ky(\xi)}V^2\right), \label{total energy}
\end{align}

\begin{figure}[t]
\begin{center}
\includegraphics[width=6cm]{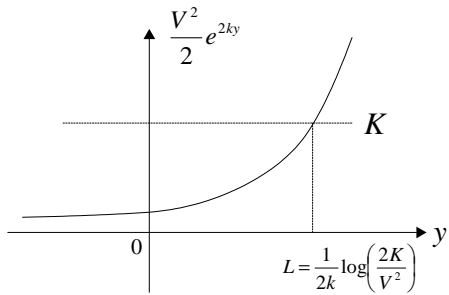}
\end{center}
\label{fig:2}
\caption{The figure shows the conservation of $K=\frac{1}{2}{y}^{\prime 2}+\frac{V^2}{2}e^{2ky}$. It is obvious that ${y}^\prime =0$ at the turning point $L=\frac{1}{2k}\log\left(\frac{2K}{V^2}\right)$; and so, we have $K=\frac{V^2}{2}e^{2kL}$ at $y=L$.}
\end{figure}

\noindent
where $x^{\prime\hat{\mu}}=\frac{dx^{\hat{\mu}}}{d\xi}$. Equation (\ref{total energy}) says that if we read $y$ as the coordinate of a particle with unite mass under the potential $\frac{1}{2}e^{2ky}V^2$, then $K$ becomes a total energy of the particle, which is related to the turning point $L$ by$K=\frac{V^2}{2}e^{2kL}$ (Fig.\ref{fig:2}).  Since $K$ is nothing but ${\cal L}$, one can obtain the expression
\begin{align}
 \sigma_{2,1}=K(\xi_2-\xi_1)^2,
\end{align}
in a single-valued region of $y$. When we write down the right-hand side of this equation by $K$ and $V^{\mu}$, we firstly be careful about the distinction of two kinds of geodesics $x_{+}^{\hat{\mu}}(\xi)$ with $y^\prime(\xi)>0,\,(\xi_0 \leq \xi \leq \xi_L)$ and  $x_{-}^{\hat{\mu}}(\xi)$ with $y^\prime(\xi)<0,\,(\xi_L \leq \xi \leq \xi_0,\,)$. Then, using the abbreviation $({x}_{i,j}^\mu)_\pm =x_\pm^\mu(\xi_i)-x_\pm^\mu(\xi_j)$ and $|{x}_{i,j}|_\pm=\sqrt{({x}_{i,j})^2_\pm}$,  we can get the following formula (Appendix A):
\begin{align}
 \sigma_{2,1} &= \frac{1}{2k^2}\left[ \tanh^{-1}\left(ke^{-kL}|{x}_{L,2}|_{\pm}\right)-\tanh^{-1}\left(ke^{-kL}|{x}_{L,1}|_{\pm}\right) \right]^2 \label{sigma-x0}
\end{align}
on condition that $ke^{-kL}|{x}_{i,j}|_\pm <1$. The (\ref{sigma-x0}) is the same for $\pm$; and, as a result, we do not have to worry about that signs. In what follows, we deal with the geodesic starting with $y=L$ along $x_{-}$ line without notice.

The potential $U(x_{(2)},x_{(1)})$ defined by (\ref{sigma-x0}) is, then, not a function of the translational invariant variable ${x}_{2,1}$ due to the curvature in the  AdS${}_5$ spacetime. Furthermore, it is not applicable for a long-distance interval $ke^{-kL}|x_{j,i}|\gg 1$. However, if we confine our attention to a case such that the two particles are located near $y=L$ brane, the low-energy IR brane, then the situation will be changed. In this case, the geodesic interval can be expressed as follows:
\begin{align}
 \sigma_{2,1}\simeq 
\begin{cases}
\frac{1}{2}e^{-2\tilde{L}}|x_{2,1}|^2 & e^{-\tilde{L}}|\tilde{x}_{2,1}| < 1 \hspace{5mm} (a) \, ,\\ 
\frac{1}{2}\tilde{L}^2e^{-2\tilde{L}}|x_{2,1}|^2  & e^{-\tilde{L}}|\tilde{x}_{2,1}| \gtrsim 1  \hspace{5mm} (b)\,,
\end{cases}  \label{sigma-Ri}
\end{align}
where the tilde denotes the scaled variables $\tilde{L}=kL,\tilde{x}=kx$, and so on.  The result implies that the geodesic interval becomes $\sim\frac
{1}{2}\tilde{L}^2$ near $ e^{-\tilde{L}}|\tilde{x}_{2,1}|\sim 1$. Since, however, we are interested in the bi-local system with an extension such as $|{x}_{2,1}| > e^{\tilde{L}}l \gg l$, we should apply (\ref{sigma-Ri}-b) to define the potential (\ref{potential}) for the present practical application. 
Thus, the two-body potential under those considerations is
\begin{align}
 U(x_{(2)},x_{(1)})=U_{2,1}=(\kappa \tilde{L})^2e^{-2\tilde{L}}|x_{2,1}|^2+\omega, \label{app-potential}
\end{align}
where $\omega$ is a free parameter effecting on the ground state mass of the bi-local system. We also stress that we may regard the $|x_{2,1}|$ in the right-hand side is independent of $y$ due to $e^{-\tilde{L}}|\tilde{x}_{2,1}| \gg 1$; and, that the resultant two-body potential is fortunately invariant under the translation of four-dimensional variables $x_{(i)}^\mu,(i=1,2)$.

\section{The wave equation of bi-local system in AdS${}_5$ spacetime}

The wave equation of the bi-local field in AdS${}_5$ spacetime is q-number representation of the constraints derived from the action (\ref{action}). The Lagrangian out of this action defines the canonical momenta $(p_{(i)},\pi_{(i)})$ conjugate to $(x_{(i)},y_{(i)})$ in the following form:
\begin{align}
 {\cal L} =\frac{1}{2}&\sum_{i=1}^2 \left[ \frac{1}{e_{(i)}}\left\{\eta_{\mu\nu}e^{-2ky_{(i)}}\dot{x}_{(i)}^\mu \dot{x}_{(i)}^\nu+\left(\dot{y}_{(i)}-(-1)^i\theta \right)^2\right\} -U_{2,1}e_{(i)} \right], \\
 p_{(i)_\mu} &=\frac{\partial{\cal L}}{\partial \dot{x}_{(i)}^\mu}=\frac{1}{e_{(i)}}e^{-2ky_{(i)}}\dot{x}_{(i)\mu}, \\
  \pi_{(i)} &=\frac{\partial{\cal L}}{\partial\dot{y}}=\frac{1}{e_{(i)}}\left(\dot{y}_{(i)}-(-1)^i\theta \right) .
\end{align}
Under the definition of those canonical momenta, the variations of the Lagrangian with respect to $e_{(i)},(i=1,2)$ and $\theta$ give rise to the constraints
\begin{align}
 H_i &\equiv -2\frac{\partial L}{\partial e_{(i)}} =e^{2ky_{(i)}}p_{(i)}^2+\left(\pi_{(i)}^2 +U_{2,1} \right)=0~~(i=1,2), \label{H_i}
\end{align}
and
\begin{align}
 \bar{\pi} &\equiv \frac{1}{2}\frac{\partial{\cal L}}{\partial \theta}=\frac{1}{2}\left(\pi_{(1)}-\pi_{(2)}\right). \hspace{43mm} \label{paibar}
\end{align}
The constraints (\ref{H_i}) and (\ref{paibar}) are not compatible each other; then, we eliminate $\bar{\pi}$ with its conjugate variable $\bar{y}=\left(y_{(1)}-y_{(2)}\right)$ strongly by means of the Dirac bracket for the second class constraints $\bar{\pi}=\bar{y}=0$. After that, we do not need to worry about the degrees of freedom $(\bar{y},\, \bar{\pi})$. Then using the combinations $\frac{1}{4}{H}=\frac{1}{2}\sum_i e^{-2ky_{(i)}}{H}_{(i)}$ and ${T}=-\frac{1}{2}\sum_i (-1)^i e^{-2ky_{(i)}}{H}_{(i)}$, the constraints (\ref{H_i}) can be written as
\begin{align}
 \frac{1}{4}{H} &\equiv \left[ \frac{1}{4}P^2+\bar{p}^2+e^{-2ky}\left(\frac{1}{4}\pi^2+U_{2,1}\right)\right]=0, \label{H=0}\\
 {T} & \equiv P\cdot\bar{p} =0 , \label{T=0}
\end{align} 
where $P=\left(p_{(1)}+p_{(2)}\right)$, $\bar{p}=\frac{1}{2}\left(p_{(1)}-p_{(2)}\right)$, and $\pi=\left(\pi_{(1)}+\pi_{(2)}\right)$ are the momenta conjugate to $X=\frac{1}{2}\left(x_{(1)}+x_{(2)}\right)$, $\bar{x}=\left(x_{(1)}-x_{(2)}\right)$, and $y=\frac{1}{2}\left(y_{(1)}+y_{(2)}\right)$, respectively.

The canonical quantization is carried out by replacing the Dirac bracket by the commutator. Then the q-number counterparts of (\ref{H=0}) and (\ref{T=0}) define respectively a master wave equation and its subsidiary condition. In the case of flat $(k=0)$ spacetime, those equations are reduced to bi-local field equations in five-dimensional Minkowski spacetime.  In such a reduced system, the condition (\ref{T=0}) is understood in the sense of expectation value by a physical state $\langle\Psi|{T}|\Psi\rangle=0$ or by ${T}^{(+)}|\Psi\rangle=0$, where ${T}^{(+)}$ is a part of ${T}$ written by the annihilation operators defined out of $(\bar{p},\bar{x})$. Then the equations $H|\Psi\rangle=0$ and ${T}^{(+)}|\Psi\rangle=0$ come to be compatible each other; and so, there arise no ghost states due to time-like oscillations of the bi-local system.

In the curved spacetime with $k\neq 0$, we can not apply this method directly to equations (\ref{H=0}) and (\ref{T=0}).  First, we have to make clear the operator ordering of $e^{-2ky}\pi^2$ in q-number theory. In what follows, we simply take the Weyl ordering
\begin{align}
 W\equiv \left(e^{-2ky}\pi^2\right)_W=e^{-2ky}(\pi+ik)^2.
\end{align}
Thus the wave equation and its subsidiary condition in q-number theory become
\begin{align}
 \left[ \frac{1}{4}P^2+\bar{p}^2+\left( \frac{1}{4}W+e^{-2ky}U_{2,1} \right) \right]\Psi=0, \label{QH=0}\\
 \left(P\cdot\bar{p}\right)^{(+)}\Psi=0, \hspace{30mm} \label{QT=0}
\end{align}
where the definition of $(P\cdot\bar{p})^{(+)}$ is not given in this stage.

The operator $W$ has the eigenstates $\phi_\lambda(z)=\frac{z}{\sqrt{N_i}}J_0\left(\frac{\sqrt{\lambda}}{k}z\right),\,(z=e^{\tilde{y}},\,N_i=\mbox{const.})$  (Appendix B) associated with the boundary condition $\left.\frac{d}{dz}\phi_{\lambda}(z)\right|_{y=L}=0$, whose roots $r_1,r_2,\cdots$ determine the eigenvalues so that $\lambda_i= \left(e^{-\tilde{L}}kr_i \right)^2,(i=1,2,\cdots)$. Then the  $\Psi$ satisfying the boundary condition $\left.\partial_y\Psi\right|_{y=L}=0$ can be expanded by a Fourier-Bessel series such as
\begin{align}
 \Psi(X,\bar{x},y)=\sum_{n=1}^\infty a_n \Phi_n(X,\bar{x})\phi_{\lambda_n}(y), \label{F-B-expansion}
\end{align}
where the coefficient $a_n$ decreases according as n increases, since $\phi_{\lambda_n}$ rapidly oscillates for a large $n$. 

Until now, the spring constant $\kappa^2$ and $\tilde{L}$ are free parameters; in what follows, we put restriction on those parameters by the conditions in UV and IR branes. First, in UV brane with $y=0$, we require  by taking $|\bar{x}|>e^{\tilde{L}}L$ into account that the order of $\bar{x}$-potential term becomes $e^{-2\tilde{L}}(\kappa\tilde{L})^2|\bar{x}|^2 > (\frac{\kappa}{k}\tilde{L}^2)^2 \gg \lambda_n \sim e^{-2\tilde{L}}k^2$; then, we obtain the first condition $\kappa \gg e^{-\tilde{L}}k^2/\tilde{L}^2$.  In this case, the order of the eigenvalues $\lambda_n$'s are negligible small compared with that of $U_{2,1}$ in UV brane even for a large $r_n$, since the $a_n$ in (\ref{F-B-expansion}) itself is decaying according as $r_n\rightarrow \infty$. Thus, we discard the $W$ term in (\ref{QH=0}) at UV brane.

\begin{figure}
\begin{center}
\includegraphics[width=6cm]{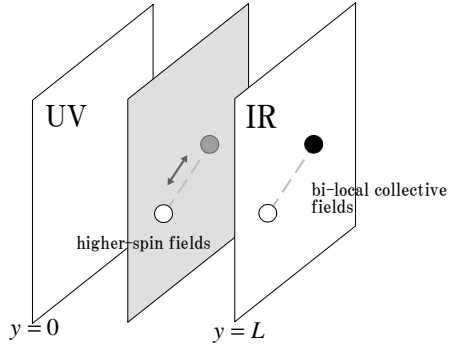}
\end{center}
\caption{The bi-local systems in respective branes}
\label{fig:3}
\end{figure}

Subsequently, we move the bi-local system from UV brane to the brane with $y>0$ (Fig.\ref{fig:3}); then, (\ref{QH=0}) and (\ref{QT=0}) take the following simple forms:
\begin{align}
 \left[ \frac{1}{4}P^2+\bar{p}^2+ \left(e^{-(\tilde{L}+\tilde{y})}\kappa \tilde{L}\right)^2\bar{x}^2 +e^{-2\tilde{y}}\omega \right] \Psi(X,\bar{x},y) =0, \label{QH1=0} \\
 \left(P\cdot\bar{p}\right)^{(+)}\Psi(X,\bar{x},y) =0. \hspace{30mm} \label{QT1=0}
\end{align}
As a matter of course, hereafter, the $y$ in those equations should be treated as a parameter instead of a dynamical variable, otherwise the bi-local system allows contiguous spectrum.

The next task is to determine the $(P\cdot\bar{p})^{(+)}$; for this purpose, we introduce $y$-dependent $\kappa_y=e^{-(\tilde{L}+\tilde{y})} \kappa\tilde{L}$. Then, we can say that $\kappa_0=e^{-\tilde{L}}\kappa\tilde{L}$ and $\kappa_L=e^{-2\tilde{L}}\kappa\tilde{L}$ are the square roots of spring constants in UV and IR branes, respectively. With this $y$-dependent $\kappa_y$, we define the $y$-dependent oscillator variables such that
\begin{align}
 \bar{x}_\mu=\sqrt{\frac{1}{2\kappa_y}}\left(a_\mu^\dag+a_\mu \right)~~\mbox{and}~~\bar{p}_\mu=i\sqrt{ \frac{\kappa_y}{2}}\left(a_\mu^\dag-a_\mu \right),
\end{align}
to which one can verify $[\bar{x}_\mu,\bar{p}_\nu]=i\eta_{\mu\nu} \Leftrightarrow [a_\mu,a_\nu^\dag]=\eta_{\mu\nu}$. In terms of those oscillator variables, (\ref{QH1=0}) can be written as
\begin{align}
 \left[ \frac{1}{4}P^2 + 2\kappa_y\left( a^\dag\cdot a+2 \right)+e^{-2\tilde{y}}\omega \right] \Psi(X,\bar{x},y) =0, \label{QH2=0}
\end{align}
from which one can say that $\alpha_y^\prime=(8\kappa_y)^{-1}$ is the Regge slope parameter in a $y$-fixed brane. Then, as the second condition on $(\kappa,\tilde{L})$, we require $\kappa_L \lesssim 10^{-20}k^2$ so as to obtain almost infinite slope parameter $\alpha_L^\prime$ at IR brane. Both conditions at UV and IR branes give rise to a possible choice such as $(\kappa,\tilde{L})\sim (k^2,50)$.

As the final step in this section, we set $(P\cdot\bar{p})^{(+)}=-i\sqrt{\frac{\kappa_y}{2}}P\cdot a$, then (\ref{QT1=0}) becomes a subsidiary condition compatible with (\ref{QH1=0}). Therefore, in what follows, we read (\ref{QT1=0}) as
\begin{align}
 P\cdot a\Psi(X,\bar{x},y)=0. \label{QT2=0}
\end{align}

\section{The bi-local fields in a brane near IR one}

Let us consider the solutions of (\ref{QH2=0}) and (\ref{QT2=0}) in each $y$-fixed brane. First, the ground state of the oscillator variables $(a_\mu,a^\dag_\mu)$ defined by $a_\mu|0\rangle=0$ can be solved as
\footnote{
The accurate representation of ground state should be $|0_y\rangle$, although we have used a simple notation $|0\rangle$. The normalization of $\langle\bar{x}|0\rangle$ in the indefinite metric formalism is given by
\[ \langle 0|0 \rangle=\int_{-\infty}^{\infty}d\xi\int_{-\infty}^{\infty}d^3x|\langle i\xi,x^i|0\rangle|^2=1 \, \] }
\begin{align}
 \langle\bar{x}|0\rangle=\left(\frac{\kappa_y}{\pi}\right)e^{- \frac{\kappa_y}{2}\bar{x}^2},  \label{ground state}
\end{align}
to which (\ref{QH2=0}) yields the mass-square eigenvalue
\begin{align}
 M_{y}(0)^2=16\kappa_y+4e^{-2\tilde{y}}\omega.
\end{align}
In this stage, we adjust $\omega$ so as to be $M_{L}(0)^2=0$; that is, we put $\omega=-4\kappa\tilde{L}$. 

\begin{figure}
\begin{center}
\includegraphics[width=5cm]{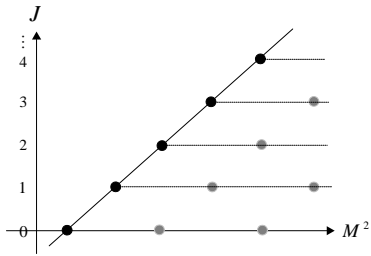}
\end{center}
\caption{The black and gray circles designate respectively  leading Regge and their daughter particles. The particles on horizontal dashed line have a common spin, whose mass will be degenerate to the ground-state one in the limit $\alpha_y^\prime\rightarrow \infty\, (y\rightarrow L)$.}
\label{fig:4}
\end{figure}

To construct the excited states of relative oscillation, one can use the physical oscillator variables $\hat{a}^\dag_\mu=\Lambda_{\mu\nu}a^{\dag\nu},\,(\Lambda_{\mu\nu}\equiv \eta_{\mu\nu}-P_\mu P_\nu/P^2)$, which tend to $(\hat{a}^{\dag\mu})=(0,a^{\dag i})$ in the rest frame $(P^\mu)=(P^0,0)$ of the bi-local system. In terms of those physical oscillator variables, one can write a complete  basis so that $\hat{a}^\dag_{\mu_1}\hat{a}^\dag_{\mu_2}\cdots \hat{a}^\dag_{\mu_J}|0\rangle,\,(J=0,1,\cdots)$. Since those states belong to reducible representations of rotation group in the rest frame of the bi-local system, it is convenient to use those states under the following combination:
\begin{align}
  |\Phi^{(m)}_{\mu_1,\cdots,\mu_J}\rangle=(\hat{a}^{\dag 2})^m \hat{a}^\dag_{(\mu_1}\hat{a}^\dag_{\mu_2}\cdots \hat{a}^\dag_{\mu_J)}|0\rangle ~ \left(m=0,1,\cdots;\, J=0,1,\cdots \right), \label{complete basis}
\end{align}
where $\hat{a}^\dag_{(\mu} \hat{a}^\dag_\nu \cdots \hat{a}^\dag_{\rho)}$ is the totally symmetric and traceless combination of $\hat{a}^\dag_\mu \hat{a}^\dag_\nu \cdots \hat{a}^\dag_\rho$; further, the state with $J=m=0$ is read as the ground state given by (\ref{ground state}).  One can verify that the state $|\Phi^{(m)}_{\mu_1,\cdots,\mu_J}\rangle$ is a simultaneous eigenstate of $N=a^\dag\cdot a$ and $Q=\hat{a}^{\dag 2}\hat{a}^{2}$ such that 
\begin{align}
 N|\Phi^{(m)}_{\mu_1\cdots\mu_J}\rangle &=(2m+J)|\Phi^{(m)}_{\mu_1\cdots\mu_J}\rangle , \\
 Q|\Phi^{(m)}_{\mu_1\cdots\mu_J}\rangle &=2m(2m+2J+1)|\Phi^{(m)}_{\mu_1\cdots\mu_J}\rangle 
\end{align}
(Appendix C); and so, the state $|\Phi^{(m)}_{\mu_1,\cdots,\mu_J}\rangle$ represents the bi-local system with mass square 
\begin{align}
 M_{y}(J,m)^2=(8\kappa_y)(J+2m)+M_y(0)^2. \label{Regge}
\end{align}
In particular, since the spin operator $\bm{S}$ of the bi-local system in the rest frame satisfies $\bm{S}^2=N(N+1)-Q$, the state $|\Phi^{(0)}_{\mu_1,\cdots,\mu_J}\rangle$ belongs to an irreducible spin representation with the highest 
spin $J$. Then (\ref{Regge}) implies that the particles represented by  $|\Phi^{(0)}_{\mu_1,\cdots,\mu_J}\rangle$ exist on a leading Regge trajectory with a slope parameter $\alpha^\prime_y\equiv (8\kappa_y)^{-1}$ (Fig.\ref{fig:4}). Thus, the general solution of (\ref{QH2=0}) and (\ref{QT2=0}) with a fixed $J$ has the form
\begin{align}
 \Phi_{y;\mu_1\cdots \mu_J}(X,\bar{x}) =\sum_{m=0}^\infty  &\int d^4P \delta^{(4)}\left(P^2+M_{y}(J,m)^2 \right)e^{iP\cdot X} \nonumber \\
 & \times C_{(m)}(P)\left(\frac{\pi}{\kappa_y}\right)\langle\bar{x}|\Phi^{(m)}_{\mu_1\mu_2\cdots\mu_J}\rangle, \label{J-states}
\end{align}
where
\begin{align}
 |\bar{x}\rangle=\left(\frac{\kappa_y}{\pi}\right) e^{-\frac{1}{2}a^\dag\cdot a^{\dag}+\sqrt{2\kappa_y}a^\dag\cdot\bar{x}}|0\rangle e^{-\frac{1}{2}\kappa_y\bar{x}^2} \label{x-eigenstate}
\end{align}
is the eigenstate such as $\bar{x}^\mu|\bar{x}^\prime\rangle=\bar{x}^{\prime\mu}|\bar{x}^\prime\rangle$ with the normalization $\langle\bar{x}^\prime|\bar{x}^{\prime\prime}\rangle=\delta^{(4)}(\bar{x}^\prime-\bar{x}^{\prime\prime})$. The factor $\left(\frac{\pi}{\kappa_y}\right)$ in the right-hand side of (\ref{J-states}) is introduced for the normalization of $\Phi_{y;\mu_1\cdots \mu_J}(X,\bar{x})$ in the limiting case of $\kappa_L\sim 0$; strictly speaking, $\kappa_L\sim 0$ means that the order of $\kappa_L$ comes to be $0$ compared with the energy scale in IR brane. It should be noticed that the states (\ref{J-states}) and (\ref{x-eigenstate}) contain the parameter $y$ through $\kappa_y$.

Now, from (\ref{x-eigenstate}), it is not difficult to evaluate
\begin{align}
 \left(\frac{\pi}{\kappa_y}\right)\langle\bar{x}|\Phi^{(m)}_{\mu_1\mu_2\cdots\mu_J}\rangle &= \left(\frac{\pi}{\kappa_y}\right)\left(\frac{\partial}{\partial k}\cdot\frac{\partial}{\partial k}\right)^m \left. \frac{\partial}{\partial k^{(\mu_1}}\cdots\frac{\partial}{\partial k^{\mu_J)}}\langle\bar{x}|e^{k\cdot \hat{a}^{\dag}}|0\rangle \right|_{k=0}  \nonumber \\
 &=\left. e^{\frac{1}{2}\kappa_y\hat{\bar{x}}^2}\left(\frac{\partial}{\partial \hat{k}}\cdot\frac{\partial}{\partial \hat{k}}\right)^m 
\frac{\partial}{\partial \hat{k}^{(\mu_1}}\cdots\frac{\partial}{\partial \hat{k}^{\mu_J)}}
e^{-\frac{1}{2}(\hat{k}-\sqrt{2\kappa_y}\hat{\bar{x}})^2}\right|_{k=0} \nonumber \\
 & \equiv e^{\frac{1}{2}\kappa_y\hat{\bar{x}}^2}S^{(m)}_{\mu_1\cdots \mu_J}\left(\sqrt{2\kappa_y}\hat{\bar{x}}\right). 
\end{align}
Since we are interested in the bi-local fields in the IR brane at $y=L$, in what follows, we consider the limiting case of $\kappa_L \sim 0$; then, one can show that $S^{(m)}_{\mu_1\cdots \mu_J}(0)=0$ for $J\neq 0$ and $S^{(m)}_{0}(0)= (-1)^m(2m+1)!!$ for $J=0$, respectively.  Therefore,  within the states (\ref{J-states}), only scalar components remain in the limit $\kappa_y \rightarrow \kappa_L \sim 0$; and then, the resultant expression to the remaining state becomes
\begin{align}
 \Phi_{L;0}(X,\bar{x}) &=\sum_{m=0}^\infty \int d\mu(P) C_{(m)}(P)S^{(m)}_{0}(0)e^{iP\cdot X} \\
 &=\sum_{m=0}^\infty \int d\mu(P) \phi^{(m)}_P(x_{(1)})\phi^{(m)}_P(x_{(2)}), \label{BL-SF}
\end{align}
where
\begin{align}
 d\mu(P)=d^4P\delta^{(4)}\left(P^2+M_{L}(0,m)^2 \right)
\end{align}
and
\begin{align}
 \phi^{(m)}_P(x_{(i)})=e^{\frac{i}{2}P\cdot x_{(i)}}\sqrt{C_{(m)}(P)S^{(m)}_0(0)}~~.
\end{align}
The $\phi^{(m)}_P(x_{(i)}), (i=1,2)$ are scalar fields associated with respective particles with the mass $\frac{1}{2}M_{L}(0,m) \sim 0$ because of $\kappa_L\sim 0$; that is, the masses of those particles are almost degenerate.  If we truncate the summation with respect to $m$ in (\ref{BL-SF}) to some number, the resultant bi-local field becomes the one, which should be compared with the bi-local collective field (\ref{BL-CF}).

\section{Summary and discussion}

In this paper, we have discussed the relation between a bi-local system embedded in AdS${}_5$ spacetime with warped metric and the higher-spin bulk fields emerging as bi-local collective fields. We tried to formulate the bi-local system in AdS${}_5$ curved spacetime in such a way that the system is reduced to two-particle bound system with a covariant harmonic oscillator potential in flat Minkowski spacetime. As a counterpart of such a harmonic oscillator potential in curved spacetime, we used the geodesic interval connecting two particles. We also modify the kinetic term of the bi-local system so that the internal relative motion is suppressed with the aid of an auxiliary gauge variable overlooking the full covariance of this formalism. 

The resultant bi-local system is characterized by three kinds of constraints. Two of them are associated with the invariance of the action under the reparametrization of time ordering parameters of respective particles; and, the other is the one due to the auxiliary gauge variable $\theta$.  In canonical formalism, the first two constraints are corresponding to on-mass-shell condition of the system and physical condition eliminating some relative motions of the bi-local system respectively.

In q-number theory, those two constraints become the wave equation of the bi-local system and its subsidiary condition, which extracts the physical states of the bi-local system, the one-particle wave function of the bi-local field. As for the constraint suppressing internal relative motion, we eliminated it beforehand as a second class constraint in the stage of classical theory. However, the remaining constraints are still not compatible each other. Then, first, we discarded the $W$ term, the kinetic term of internal center of mass variable $y$, in the UV brane; then, $y$ becomes simply a parameter designating each brane, on which the bi-local system is placed.  We further treated the $T$ condition suppressing timelike oscillations of the bi-local system in a form of expectation value; and then, the wave equation and it subsidiary condition, $T$ condition, become compatible each other.

The on-mass-shell solutions of resultant wave equation represent the particles overlying on Regge trajectories with the slope parameter $\alpha^\prime_y=(8\kappa_y)^{-1}$. Here, the $\kappa_y^2$ is a spring constant in a $y$-fixed brane, which tend to $0$ according as $y$ comes close to $L$, the place of IR brane. Strictly speaking $\sqrt{\kappa_L}$ is almost $0$ compared with the energy scale of IR brane. To realize these setup on the bi-local system, we have chosen the parameters in our model so that $(\kappa,\tilde{L})\sim (k^2,50)$, which derive the reasonable order of $\sqrt{\kappa_y}$ such as $(\sqrt{\kappa_0}, \sqrt{\kappa_L})\sim k(10^{-10},10^{-21})$ for $k\sim$ Planck energy scale.

Hence, in the IR brane, all particles degenerate in almost massless one's; furthermore, non-zero spin components of the bi-local field can be shown to vanish naturally on that brane. Therefore, the bi-local field on the IR brane behaves as the bi-local collective field (\ref{BL-CF}) out of higher-spin bulk fields as we wanted to show. It should be, however, noticed that the respective particles described by the bi-local field (\ref{BL-SF}) hold a common center of mass momentum as their hysteresis of a bound system in bulk. 

Further, from the bi-local field (\ref{BL-SF}), we cannot say anything about 1) the bound system of particles laid on different branes and 2) the bi-local system with very small interval such as $|x_{2,1}| \lesssim e^{-\tilde{L}}l$.  In relation with 2), we should also notice that the practical interaction between two particles on the IR brane can take place only under discreet distances with the unit of $ e^{-\tilde{L}}l$. This property of the interaction evokes the behavior of elementary domains proposed by Yukawa\cite{domain-1,domain-2}. Those are important and interesting future problems to make clear the relation between the bi-local system and the bi-local collective field.

\section*{Acknowledgment}

The authors wish to thank the members of the theoretical group in Nihon University for their interest in this work and comments. We are thankful to Dr. N. Kanda and Dr. T. Takanashi for their stimulating discussions. One of us (S. N.\,) also would like to thank to Mr. R. Satake for the discussion in the early stage of this work
\footnote{
In flat Minkowski spacetime, the bi-local collective field (\ref{BL-CF}) has a similar structure to the three vertex function of bi-local fields\cite{Vertex}. Indeed, R. Satake\cite{Satake-1} deduced that (\ref{BL-CF}) could be understood as an infinite slope limit of such a vertex function. After completing this paper, we found \cite{Satake-2}, which discuss the same line of approach as \cite{Satake-1}. }

\appendix

\section{The geodesic in AdS${}_5$ within $0<y(\xi)<L$.}

Regarding ${\cal L}$ as the Lagrangian of generalized coordinates $(x^\mu(\xi),y(\xi))$ in AdS${}_5$ described by (\ref{metric}), the equations (\ref{constant velocity}) and (\ref{total energy}) are respectively direct results of equations of motion with respect to $x^\mu(\xi)$ and $y(\xi)$. As can be seen from Fig.5, the $y(\xi)$ is a multi-valued function of $\xi$. Changing the variable $\zeta(\xi)=e^{-ky(\xi)}$,  (\ref{constant velocity}) can be rewritten as
\begin{align}
 \zeta^\prime=\pm k\sqrt{2K}\sqrt{\zeta^2-V^2/(2K)},
\end{align}
which can be integrated easily in a single valued region of $y$ so that
\begin{align}
 \zeta_\pm(\xi) =e^{-k y_\pm(\xi)}=\sqrt{\frac{V^2}{2K}}\cosh\left\{\mp k\sqrt{2K}(\xi-\xi_*)+c_*  \right\}, \label{z-solution}
\end{align}
where the $\pm$ in (\ref{z-solution}) designate the sign of $y^\prime$; the $c_*$ is a constant related to one end of $\gamma$ such as $y_*=y(\xi_*)$. In what follows, we choose simply $y_*=L$ ($\xi_*=\xi_L$) one turning point of $\gamma$, which leads to $c_*=0$ because of $e^{-kL}=\sqrt{\frac{V^2}{2K}}$ as pointed out in Fig.\ref{fig:2}. In this case, $x_{+}^\mu(\xi)$ and $x_{-}(\xi)$ become functions defined respectively in the region $\xi_0 \leq \xi \leq \xi_L$ and $\xi_L \leq \xi \leq \xi_0$, where $\xi_0$ and $\xi_L$ are points such as by $y(\xi_0)=0$ and $y(\xi_L)=L$. Then, substituting (\ref{z-solution}) in this case for (\ref{constant velocity}), we can integrate $x_\pm^{\prime\mu}(\xi)$ as
\begin{align}
 x_\pm^\mu(\xi)-x_\pm^\mu(\xi_L)=\hat{V}^\mu\frac{e^{kL}}{k}\tanh\left\{k\sqrt{2K}(\xi-\xi_L) \right\} \label{x-line}
\end{align}
where $\hat{V}^\mu=V^\mu/\sqrt{V^2}$ is a constant unit vector for the direction of $x^{\prime\mu}$. Hereafter, we write $x^\mu_{b,a}\equiv x^\mu(\xi_b)-x^\mu(\xi_a)$, which allows to express $|x_{b,a}|\equiv \sqrt{(x_{b,a})^2}=(x_{b,a})\cdot\hat{V}$ for $\xi_a <\xi_b$.  It is also convenient to use the symbol $\tilde{A}=kA$, which is the $A$ measured by $k^{-1}\sim l$.  Then (\ref{x-line}) can be written as
\begin{align}
 |\tilde{x}_{i,L}|_{\pm}=e^{\tilde{L}}\tanh\left\{\sqrt{2K}(\tilde{\xi}_{i,L})\right\}
,~\left(\,\tilde{\xi}_{i,j}=k(\xi_i-\xi_j)\, \right) ,
\end{align}
from which the following follows
\begin{align}
 \tilde{\xi}_{b,a}=\frac{1}{\sqrt{2K}}\left[ \tanh^{-1}\left(e^{-\tilde{L}}|\tilde{x}_{b,L}|\right)_{\pm}-\tanh^{-1}\left(e^{-\tilde{L}}|\tilde{x}_{a,L}|\right)_{\pm} \right] \label{diff-1}
\end{align}
providing $e^{-\tilde{L}}|\tilde{x}_{i,L}|_{\pm} <1,\,(i=a,b)$. The result means that we do not need to worry about the $\pm$ when we represent  $\tilde{\xi}_{b,a}$ in terms of $\tilde{x}_{i,L}$. Further, (\ref{z-solution}) yields another expression to $\tilde{\xi}_{b,a}\,(>0)$ such that
\begin{align}
 \tilde{\xi}_{b,a}=\frac{\mp}{\sqrt{2K}}\left[\cosh^{-1}\left(e^{-\tilde{y}_{b,L}}\right)_\pm -\cosh^{-1}\left(e^{-\tilde{y}_{a,L}}\right)_\pm \right], \label{diff-2}
\end{align}
which gives rise to
 \begin{align}
 \sigma_{0,L}=K\xi_{0,L}^2=\frac{1}{2k^2}\left\{\cosh^{-1}\left(e^{\tilde{L}}\right)\right\}^2\simeq \frac{L^2}{2}. \label{sigma-y}
\end{align}
If we apply (\ref{diff-1}) formally to $|\tilde{\xi}_{0,L}|$, then $\sigma_{0,L} \simeq L^2/2$ requires $e^{-\tilde{L}}|\tilde{x}_{0,L}|=\tanh(\tilde{L})\sim 1$. Since, however, $e^{-\tilde{L}}|\tilde{x}_{0,L}|\sim 1$ is near the applicable limit of (\ref{diff-1}), we must be careful to evaluate it.  The right value of $|\tilde{x}_{0,L}|$ can be obtained from 
\begin{align}
 |\tilde{x}_{i,L}|=e^{\tilde{L}}\tanh\left\{\cosh^{-1}\left(e^{-\tilde{y}_{i,L}}\right)\right\}=e^{\tilde{L}}\sqrt{1-e^{2\tilde{y}_{i,L}}}\, , \label{x(y)}
\end{align}
which is obtained by substituting (\ref{diff-2}) for (\ref{x-line}). From this equation, one can find that $|\tilde{x}_{i,L}|$ runs from $|\tilde{x}_{L,L}|=0$ to $|\tilde{x}_{0,L}|=e^{\tilde{L}}\sqrt{1-e^{-2\tilde{L}}}\,(\lesssim e^{\tilde{L}})$ according as $\tilde{y}_i$ runs from $\tilde{L}$ to $0$; and so, the domain of $|\tilde{x}|$ is very large against the one of $\tilde{y}$. 

Now, let us consider the case such that $\tilde{y}_i,(i=a,b)$ are located very close to $\tilde{L}$ in (\ref{diff-1}); then, it can be verified easily that
\begin{align}
 \tilde{\xi}_{b,a} \simeq \frac{1}{\sqrt{2K}}e^{-\tilde{L}}\left(|\tilde{x}_{b,L}|-|\tilde{x}_{a,L}|\right)= \frac{1}{\sqrt{2K}}e^{-\tilde{L}}|\tilde{x}_{b,a}|
\rightarrow \sigma_{b,a}\simeq \frac{1}{2}e^{-2\tilde{L}}|x_{b,a}|^2.
 \label{sigma-x}
\end{align}
\begin{figure}
\centering\includegraphics[width=7cm]{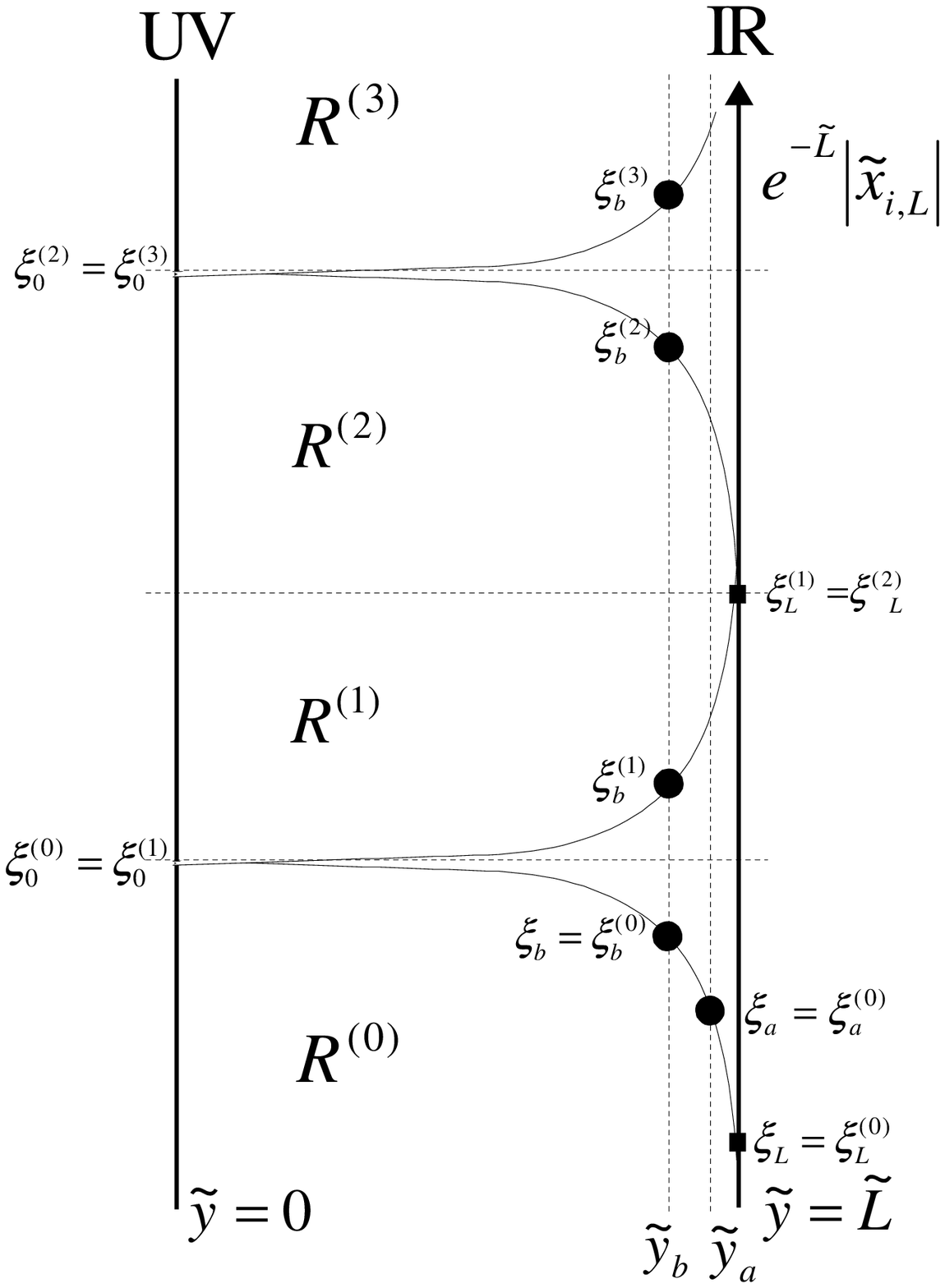}
\label{fig:5}
\caption{The symbolic figure of turning geodesic. The superscript $(n)$ in $\xi^{(n)},y^{(n)}$ designate that they characterize the geodesic in $R^{(n)}$ region in the figure.}
\end{figure}
To extent this relation to multi-valued regions of $y(\xi)$, we have to take the successive turnings of geodesic at $y=0$ and $y=L$ branes in (Fig.5) into account. Writing, here, $\xi^{(k)}_{i,j}=\xi_i^{(k)}-\xi_j^{(k)}$ and $\xi_{b^{(n)},a^{(m)}}=\xi_b^{(n)}-\xi_a^{(m)}$, it can be verified from (Fig.5) that
\begin{align}
 \tilde{\xi}_{b^{(n)},a^{(0)}}=\tilde{\xi}^{(n)}_{b.L}+\left(\tilde{\xi}^{(n)}_{L,0}+\tilde{\xi}^{(n-1)}_{0,L}+\tilde{\xi}^{(n-2)}_{L,0}+\cdots+\tilde{\xi}^{(0)}_{0,L}\right)+\tilde{\xi}^{(0)}_{L,a}
\end{align}
for a odd number of $n$; and,
\begin{align}
 \tilde{\xi}_{b^{(n)},a^{(0)}}=\tilde{\xi}^{(n)}_{b.L}+\left(\tilde{\xi}^{(n-1)}_{L,0}+\tilde{\xi}^{(n-2)}_{0,L}+\cdots+\tilde{\xi}^{(0)}_{0,L}\right)+\tilde{\xi}^{(0)}_{L,a}
\end{align}
for a even number of $n$. Furthermore, since $\tilde{\xi}^{(2n+1)}_{L,0}=\tilde{\xi}^{(2n)}_{0,L}=\frac{1}{\sqrt{2K}}\tilde{L}\,(n=0,1,\cdots)$ and
\begin{align}
 \tilde{\xi}^{(n)}_{b,L}+\tilde{\xi}^{(0)}_{L,a} \simeq \frac{1}{\sqrt{2K}}e^{-\tilde{L}}\left\{(\tilde{x}^{(n)}_b-\tilde{x}^{(n)}_L)\cdot\hat{V}+ (\tilde{x}^{(0)}_L-\tilde{x}^{(0)}_a)\cdot\hat{V} \right\}\,,
\end{align}
we are able to get the following expression:
\begin{align}
 \tilde{\xi}_{b^{(n)},a^{(0)}} \simeq \frac{1}{\sqrt{2K}}\left\{2\left[\frac{n+1}{2}\right](\tilde{L}-1)+e^{-\tilde{L}}|\tilde{x}_{b^{(n)},a^{(0)}}| \right\},
\end{align}
where $[n]$ is the largest integer being not greater than $n$. In this paper, we are interested in the bi-local bound system such as $n \gg \tilde{L}$; then, one can evaluate $2\left[\frac{n+1}{2}\right]\simeq n\simeq e^{-\tilde{L}}|\tilde{x}_{b^{(n)},a^{(0)}}|$. Therefore, under those approximations, the geodesic interval extended to $n$ turned regions becomes
\begin{align}
 \sigma(x_{b^{(n)}},x_{a^{(0)}})=\frac{1}{k^2}K\left(\tilde{\xi}_{b^{(n)},a^{(0)}}\right)^2=\frac{1}{2}\tilde{L}^2e^{-2\tilde{L}}|{x}_{b^{(n)},a^{(0)}}|^2 , \label{sigma-n}
\end{align}
in which the discrete indices $(n)$ and $(0)$ are no longer important to attach. Here, if we try to apply  (\ref{sigma-n}) to $b^{(n)}=0^{(1)}$ and $a^{(0)}=L^{(0)}$, then we can get $\sigma_{0,L}=\frac{1}{2}L^2$, the result of (\ref{sigma-y}). This implies that the$|x_{b,a}|$ in  (\ref{sigma-n}) is applicable from $|x_{b,a}| \gtrsim e^{\tilde{L}}l$ to infinity. On the other side, (\ref{sigma-x}) is holds in the $n=0$ single-valued region with $|\tilde{x}_{i,L}|\ll 1,(i=a,b)$.

\section{Eigenvalue problem of $W$}

The eigenvalue problem of $W$ can be solved easily by using the variable $z(y)=e^{ky}$. Then, by taking $\frac{d}{dy}=kz\frac{d}{dz}$ into account, the eigenvalue equation of $W$ can be written as
\begin{align}
 W\phi_\lambda(z(y)) &=e^{-2ky}\left(\pi+i k \right)^2\phi_\lambda(z(y))~~\left(\pi=-i\frac{d}{dy} \right) \\
 &=-k^2\left(\frac{d^2}{dz^2}-\frac{1}{z}\frac{d}{dz}+\frac{1}{z^2}\right)\phi_\lambda(z)=\lambda \phi_\lambda(z)
\end{align}
; that is, 
\begin{align}
 \frac{d^2\phi_\lambda}{dz^2}-\frac{1}{z}\frac{d\phi_\lambda}{dz}+\left\{ \frac{\lambda}{k^2}+\frac{1}{z^2}\right\}\phi_\lambda=0 \, . \label{phi-lambda}
\end{align}
The solutions of this equation are reduced to Bessel functions multiplied by a power of $z$; indeed, it is known that the equation
\begin{align}
 \frac{d^2 w}{dz^2}+\frac{1-2\alpha}{z}\frac{dw}{dz}+\left(\beta^2+\frac{\alpha^2-\nu^2}{z^2}\right)w=0
\end{align}
has $z^\alpha Z_\nu(\beta z)$ as its solution \cite{Bessel}, where $Z_\nu(x)$ is one of $J_\nu(x),Y_\nu(x),H_\nu^{(1)}(x)$, and $H_\nu^{(2)}(x)$.
The $\phi_\lambda$ in (\ref{phi-lambda}) is the case of $\alpha=1, \beta=\frac{\sqrt{\lambda}}{k}$, and $\nu=0$; and so, we can set
\begin{align}
 \phi_\lambda(z)= zZ_0\left(\frac{\sqrt{\lambda}}{k}z \right) .
\end{align}
Further, it is not difficult to derive from (\ref{phi-lambda}) that
\begin{align}
 \left\{\frac{1}{z}\left(\phi_{\lambda_i}\phi_{\lambda_j}^\prime - \phi_{\lambda_i}^\prime\phi_{\lambda_j}\right)\right\}^\prime =(\lambda_i-\lambda_j)\frac{1}{k^2}\frac{1}{z}\phi_{\lambda_i}\phi_{\lambda_j}, \label{total derivative}
\end{align}
where the \lq\lq prime\rq\rq denotes the derivative with respect to $z$. This means that under the boundary conditions:
\begin{align}
 \left. \frac{1}{z}\left(\phi_{\lambda_i}\phi_{\lambda_j}^\prime - \phi_{\lambda_i}^\prime\phi_{\lambda_j}\right)\right|_{z=z_0,z_L}=0
 \label{boundary}
\end{align}
with $z_0=z(0)=1$ and $z_L=z(L)=e^{\tilde{L}}$, we can put the normalization
\begin{align}
 \langle \phi_{\lambda_i}|\phi_{\lambda_j}\rangle &=\int_{z_0}^{z_L}\frac{dz}{z}\phi_{\lambda_i}(z)\phi_{\lambda_j}(z) \\
 &= \int_0^L kdy Z_{0}\left(\frac{\sqrt{\lambda_i}}{k}e^{ky}\right)Z_{0}\left(\frac{\sqrt{\lambda_j}}{k}e^{ky}\right) \propto \delta_{i,j} \, . \label{orthonormal}
\end{align}
To realize the solutions satisfying the boundary conditions, first, we take $Z_0(x)=J_0(x)$, which is finite on real $x$ line. Secondly, we require
\begin{align}
 \left. \phi_\lambda^\prime(z) \right|_{z_L}=\left. \left\{J_0\left(\frac{\sqrt{\lambda}}{k}z\right)-\frac{\sqrt{\lambda}}{k}zJ_1 \left(\frac{\sqrt{\lambda}}{k}z\right) \right\} \right|_{z_L}=0,
 \label{B.C.}
\end{align}
where $J^\prime_0(x)=-J_1(x)$; then, the $r\equiv \frac{\sqrt{\lambda}}{k}z_L$ is determined as a root of this equation. Writing the $i$-th root of (\ref{B.C.}) by $r_i$, the $i$-th eigenvalue $\lambda_i,~(i=1,2,\cdots)$ has the expression
\begin{align}
 \lambda_i = \left(\frac{k}{z_L}r_i\right)^2~(i=1,2,\cdots).
\end{align}
We note that the (\ref{B.C.}) implies not the vanishing internal wave function at $z=z_L$ but the vanishing of flux for $y \geq L$. Now, using $r_k=\frac{\sqrt{\lambda_k}}{k}z_L$, the $z^{-1}\left(\phi_{\lambda_i}\phi_{\lambda_j}^\prime - \phi_{\lambda_i}^\prime\phi_{\lambda_j}\right)$ in the left-hand side of (\ref{boundary}) at $z\,(=e^{\tilde{y}})$ can be rewritten as
\begin{align}
 \frac{z}{z_L}\left\{r_i J_1\left(r_i \frac{z}{z_L}\right)J_0\left(r_j \frac{z}{z_L}\right)-r_j J_1\left(r_j \frac{z}{z_L}\right)J_0\left(r_i \frac{z}{z_L} \right)\right\},
\end{align}
which tend to $0$ according as $z\rightarrow 1\,(\mbox{i.e.}\,\tilde{y}\rightarrow 0)$ because of
\footnote{
In ascending order of $r_i,(i=1,2,\cdots)$, the roots of (\ref{B.C.}) are obtained numerically as $r_1=1.2557\cdots,r_2=4.0794\cdots,r_3=7.1557\cdots$, and so on. On the other side, for a sufficiently large $r_i$, the (\ref{B.C.}) gives $J_1(r_i)=\frac{1}{r_i}J_0(r_i)\simeq 0$; and so, $r_i$ becomes approximately a zero $( =j_{1,i}\simeq \pi i)$ of $J_1$. This means that the condition $r_ie^{-\tilde{L}}\simeq 0$ is satisfied even for a huge number $r_M\sim e^{\frac{1}{2}\tilde{L}}$, to which  the coefficient $a_{M}$ in the expansion (\ref{F.B.}), however, tends to vanish because of rapid oscillation of $\phi_{\lambda_M}$.
}
 $J_1(r_ie^{-\tilde{L}})\simeq J_1(0)=0$. In this sense, the boundary conditions (\ref{boundary}) are satisfied at $z_0$ and $z_L$. 

In order to give the normalization of $\phi_\lambda(z)$, let us consider the limit $\lambda_i=\lambda_j+\epsilon \,(\epsilon \rightarrow 0)$ in $\langle\phi_{\lambda_i}|\phi_{\lambda_j}\rangle$  by taking (\ref{total derivative}) into account. Then, a little calculation leads to
\begin{align}
 \langle\phi_{\lambda_j}|\phi_{\lambda_j}\rangle &=\lim_{\epsilon\rightarrow 0}\frac{k^2}{\epsilon}\left[\frac{1}{z}\left\{ \phi_{\lambda_j+\epsilon}(z)\phi^\prime_{\lambda_j}(z) - \phi_{\lambda_j+\epsilon}(z)^\prime \phi_{\lambda_j}(z) \right\}\right]_{z_0}^{z_L} \\
 &=k^2\left[\frac{1}{z}\left\{
(\partial_\lambda \phi_{\lambda})(z)\phi^\prime_{\lambda}(z)-(\partial_\lambda \phi^\prime_{\lambda})(z)\phi_{\lambda}(z)~\right\}_{\lambda=\lambda_j}\right]_{z_0}^{z_L} \\
 &=\frac{z_L^2}{2}\left(1+\frac{1}{r_j^2}\right)J_0\left(r_j\right)^2~(\, \equiv N_j \,).
\end{align}
Therefore, under the rescale $\phi_{\lambda_i}(z)=\frac{z}{\sqrt{N_i}}J_0\left(\frac{\sqrt{\lambda_i}}{k}z\right)$, the eigenfunctions $\{\phi_{\lambda_i}\}$ form an independent orthonormal basis, by which any internal wave function $\Phi(y)$ can be expanded in the following series:
\begin{align}
 \Phi(y)=\sum_i a_i\phi_{\lambda_i}(y)~~(\, a_i=\langle\phi_{\lambda_i}|\Phi\rangle \,). \label{F.B.}
\end{align}  

\section{Spin eigenstates}

In the rest frame of the bi-local system with $P=(M,0,0,0)$, the hatted oscillator variables are reduced to $(\hat{a}_\mu)=(0, a_1,a_2,a_3)$ and $(\hat{a}^\dag_\mu)=(0, a^\dag_1,a^\dag_2,a^\dag_3)$. In terms of those reduced oscillator variables, the spin operator of the bi-local system is defined by $S_i=-i\epsilon_{ijk}a_j^\dag a_k,~(i,j,k=1,2,3)$, to which by taking $\epsilon_{ijk}\epsilon_{ilm}=(\delta_{jl}\delta_{km}-\delta_{kl}\delta_{jm})$ into account, one can verify
\begin{align}
S^2 &=(-i)^2\epsilon_{ijk}\epsilon_{ilm}a_j^\dag a_k a_l^\dag a_m=N(N+1)-Q~,
\end{align}
where $N=\bm{a}^\dag\cdot \bm{a}$ and $Q=\bm{a}^{\dag 2}\bm{a}^2$. Since $N$ and $Q$ are commute each other, there are common eigenstates of those operators. In particular the spin eigenstates with zero eigenvalue of $Q$ have the following form:
\begin{align}
 a^\dag_{(i_1}a^\dag_{i_2}\cdots a^\dag_{i_J)}|0\rangle &\equiv  \left[a^\dag_{i_1}a^\dag_{i_2}\cdots a^\dag_{i_J}-\left(d+\begin{pmatrix}J\\2\end{pmatrix}-1\right)^{-1}\sum_{(a,b)}\delta_{i_ai_b} \bm{a}^{\dag 2} a^\dag_{i_1}\cdots \widehat{a^\dag_{i_a}}\cdots \widehat{a^\dag_{i_b}}\cdots a^\dag_{i_J}\right]|0\rangle \nonumber \\
 &=T_{i_1\cdots i_J}^{j_i\cdots j_J}a^\dag_{j_1}\cdots a^\dag_{j_J}|0\rangle ~~~(d=\delta_i^i=3), 
\end{align}
where $\sum_{(a,b)}$ stands for the summation over two indices $(a,b)$ taken at a time out of $J$ different objects $(1,2,\cdots,J)$; and, 
\begin{align}
T_{i_1\cdots i_J}^{j_i\cdots j_J}=\delta_{i_1}^{j_1}\cdots\delta_{i_J}^{j_J}-\left(d+\begin{pmatrix}J\\2\end{pmatrix}-1 \right)^{-1} \sum_{(a,b)}\delta_{i_ai_b}\delta^{j_aj_b} \delta_{i_1}^{j_1}\cdots \widehat{\delta_{i_a}^{j_a}} \cdots \widehat{\delta_{i_b}^{j_b}} \cdots \delta_{i_J}^{j_J}.
\end{align}
One can see that the $T_{i_1\cdots i_J}^{j_i\cdots j_J}$ is symmetric and traceless with respect to both of the subscripted indices $(i_1,\cdots,i_J)$ and the superscripted indices $(j_1,\cdots,j_J)$; then, it is not difficult to verify that 
\begin{align}
 Q a^\dag_{(i_1}a^\dag_{i_2}\cdots a^\dag_{i_J)}|0\rangle &=0 , \\
 \bm{S}^2a^\dag_{(i_1}a^\dag_{i_2}\cdots a^\dag_{i_J)}|0\rangle &=J(J+1)a^\dag_{(i_1}a^\dag_{i_2}\cdots a^\dag_{i_J)}|0\rangle .
\end{align}
Thus the states $\{a^\dag_{(i_1}a^\dag_{i_2}\cdots a^\dag_{i_J)}|0\rangle\}$ form a spin-$J$ irreducible representation of $O(3)$ group; and, $N$ becomes the highest spin operator in that representation. Meanwhile, the state $|\Phi^{(m)}_{(i_1\cdots i_J)}\rangle=(\bm{a}^{\dag 2})^m a^\dag_{(i_1}a^\dag_{i_2}\cdots a^\dag_{i_J)}|0\rangle\,(m\neq 0)$ is not a zero eigenstate of $Q$, although it again satisfies
\begin{align}
 \bm{S}^2(\bm{a}^{\dag 2})^m |\Phi^{(m)}_{(i_1\cdots i_J)}\rangle =J(J+1)(\bm{a}^{\dag 2})^m |\Phi^{(m)}_{(i_1\cdots i_J)}\rangle \,. 
\end{align}
The eigenvalue of $Q$ can be found so that
\begin{align}
 Q|\Phi^{(m)}_{(i_1\cdots i_J)}\rangle &=\bm{a}^{\dag 2}\bm{a}^{2}\bm{a}^{\dag 2}(\bm{a}^{\dag 2})^{m-1}a^\dag_{(i_1}a^\dag_{i_2}\cdots a^\dag_{i_J)}|0\rangle \nonumber \\
 &=\bm{a}^{\dag 2}(6+4\bm{a}^\dag\cdot\bm{a}+\bm{a}^{\dag 2}\bm{a}^2)(\bm{a}^{\dag 2})^{m-1}a^\dag_{(i_1}a^\dag_{i_2}\cdots a^\dag_{i_J)}|0\rangle \nonumber \\
&=\bm{a}^{\dag 2}\left[ Q+(4J+6) +8(m-1)\right] (\bm{a}^{\dag 2})^{m-1}a^\dag_{(i_1}a^\dag_{i_2}\cdots a^\dag_{i_J)}|0\rangle \nonumber \\
 &=(\bm{a}^{\dag 2})^2\left[ Q+2(4J+6) +8\{(m-1)+(m-2)\}\right] (\bm{a}^{\dag 2})^{m-2}a^\dag_{(i_1}a^\dag_{i_2}\cdots a^\dag_{i_J)}|0\rangle \nonumber \\
 &\vdots \nonumber \\
& =2m(2J+2m+1)|\Phi^{(m)}_{(i_1\cdots i_J)}\rangle\, .
\end{align}

Now the $x$ representation of $|\Phi^{(m)}_{(i_1\cdots i_J)}\rangle$ can be evaluated by the expression
\begin{align}
 \left(\frac{\pi}{\kappa_y}\right)\langle\bar{\bm{x}}|\Phi^{(m)}_{i_1\cdots i_J}\rangle =\left. e^{\frac{1}{2}\kappa_y \bar{\bm{x}}^2}\left(\frac{\partial}{\partial {\bm k}}\cdot\frac{\partial}{\partial {\bm k}}\right)^m 
T_{i_1\cdots i_J}^{j_1\cdots j_J}\frac{\partial}{\partial k^{j_1}}\cdots\frac{\partial}{\partial k^{j_J}}
e^{-\frac{1}{2}(\bm{k}-\sqrt{2\kappa_y}\bar{\bm{x}})^2}\right|_{\bm{k}=0} \label{spin representation}
\end{align}
The case of $J=0$ is particularly simple, and we obtain
\begin{align}
 \left(\frac{\pi}{\kappa_y}\right)\langle\bar{x}|\Phi^{(m)}_0\rangle &=\left. e^{\frac{1}{2}\kappa_y \bar{\bm{x}}^2}\left(\frac{\partial}{\partial t}\right)^m e^{t\left(\frac{\partial}{\partial\bm{k}}\right)^2}e^{-\frac{1}{2}(\bm{k}-\sqrt{2\kappa_y}\bar{\bm{x}})^2}\right|_{\bm{k}=0,t=0} \nonumber \\
 &=\left. e^{\frac{1}{2}\kappa_y \bar{\bm{x}}^2}\left(\frac{\partial}{\partial t}\right)^m \int \frac{d^3\bm{z}}{\sqrt{(2\pi)^3}}e^{-\frac{1}{2}\bm{z}^2+\sqrt{2t}\bm{z}\cdot\left(\frac{\partial}{\partial\bm{k}}\right) }
e^{-\frac{1}{2}(\bm{k}-\sqrt{2\kappa_y}\bar{\bm{x}})^2}\right|_{\bm{k}=0,t=0} \nonumber \\
 &=\left. e^{-\frac{1}{2}\kappa_y \bar{\bm{x}}^2}\left(\frac{\partial}{\partial t}\right)^m \frac{1}{\sqrt{(1+2t)^3}}e^{-2\frac{t}{1+2t}\kappa_y\bar{\bm{x}}^2}\right|_{t=0} \\
 &\rightarrow (-1)^m(2m+1)!!~~(\kappa_y\rightarrow 0).
\end{align}
The limit in the last expression is taken, since we are interested in the result in the IR brane at $y=L$, in which the $\kappa_y$ is virtually zero. On the same footing, let us consider the case of $J\ne 0$ in the limit $\kappa_y=0$. Then (\ref{spin representation}) becomes
\begin{align}
 \left(\frac{\pi}{\kappa_y}\right)\langle\bar{\bm{x}}|\Phi^{(m)}_{i_1\cdots i_J}\rangle &=\left. \left(\frac{\partial}{\partial \bm{k}}\cdot\frac{\partial}{\partial \bm{k}}\right)^m 
T_{i_1\cdots i_J}^{j_1\cdots j_J}\frac{\partial}{\partial k^{j_1}}\cdots\frac{\partial}{\partial k^{j_J}}
e^{-\frac{1}{2}\bm{k}^2}\right|_{\bm{k}=0} \label{spin-J 1}\\
 &=\left. \left\{\left(\frac{\partial}{\partial \bm{k}}-\bm{k}\right)^2\right\}^m 
T_{i_1\cdots i_J}^{j_1\cdots j_J}
\left(\frac{\partial}{\partial k^{j_1}}-k_{j_1}\right)\cdots\left(\frac{\partial}{\partial k^{j_J}}-k_{j_J}\right)\right|_{\bm{k}=0} \label{spin-J 2}
\end{align}
The right-hand side vanishes for an odd number of $J$, since the (\ref{spin-J 1}) is an odd function of $\bm{k}$. For an even number of $J$, the non-vanishing terms at $\bm{k}=0$ are consisting of the terms such as $\frac{\partial}{\partial k^{j_a}}k_{i_b}=\delta_{j_aj_b}$ and $\left(\frac{\partial}{\partial k^l}k_{j_a}\right)\left(\frac{\partial}{\partial k^l}k_{j_b}\right)=\delta_{lj_a}\delta_{lj_b}=\delta_{j_aj_b}$. Those terms, however, again give rise to vanishing contribution due to the traceless property of $T_{i_1\cdots i_J}^{j_1\cdots j_J}$. Therefore, the state (\ref{spin-J 2}) vanishes in the IR brane.

\end{document}